\pdfoutput=1 
\documentclass[12pt]{iopart}
\usepackage{iopams}  
\usepackage{bm}
\usepackage{graphicx}
\begin{document}

\newcommand{\unit}[1]{\ensuremath{~\mathrm{#1}}}
\newcommand{\unitnospace}[1]{\ensuremath{\mathrm{#1}}}
\newcommand{\K}{\unit{K}}
\newcommand{\T}{\unit{T}}
\newcommand{\mT}{\unit{mT}}
\newcommand{\uT}{\unit{\mu{}T}}
\newcommand{\A}{\unit{A}}
\newcommand{\nA}{\unit{nA}}
\newcommand{\pA}{\unit{pA}}
\newcommand{\kV}{\unit{kV}}
\newcommand{\uV}{\unit{\mu{}V}}
\newcommand{\mm}{\unit{mm}}
\newcommand{\um}{\unit{\mu{}m}}
\newcommand{\nm}{\unit{nm}}
\newcommand{\degree}{\unitnospace{^\circ}}

\newcommand{\E}[1]{\ensuremath{\!\times\! 10 ^{#1}}}

\newcommand{\vs}{vs\xspace}						
\newcommand{\ie}{i.e.\ }							

\SUST

\title[Theory and experiment testing flux-line-cutting physics]{Theory and experiment testing flux-line-cutting physics}

\author{John R Clem,$^1$\footnote[3]{To whom correspondence should be
addressed.} Marcus Weigand,$^2$ J H Durrell,$^3$ and A M Campbell$^3$}

\address{$^1$\ Ames Laboratory and Department of Physics and Astronomy,\\
  Iowa State University, Ames, Iowa, 50011--3160, USA}

\address{$^2$\ Department of Materials Science and Metallurgy, University of Cambridge,
Pembroke Street, Cambridge, CB2 3QZ, UK}

\address{$^3$\ Department of Engineering,  University of Cambridge, Trumpington Street, Cambridge CB2 1PZ, UK}

 \ead{clem@ameslab.gov}

\begin{abstract}We discuss predictions of five proposed theories for the critical state of type-II superconductors accounting for both flux cutting and flux transport (depinning).  The theories predict different behaviours for the ratio $E_y/E_z$ of the transverse and parallel components of the in-plane electric field produced just above the critical current of a type-II superconducting slab as a function of the angle of an in-plane applied magnetic field.  We present experimental results measured using an epitaxially grown YBCO thin film favoring one of the five theories: the extended elliptic critical-state model.  We conclude that when the current density $\bm J$ is neither parallel nor perpendicular to the local magnetic flux density $\bm B$, both flux cutting and flux transport occur simultaneously when $J$ exceeds the critical current density $J_c$, indicating an intimate relationship between flux cutting and depinning.  We also conclude that the dynamical properties of the superconductor when $J$ exceeds $J_c$ depend in detail upon two nonlinear effective resistivities for flux cutting ($\rho_c$) and flux flow ($\rho_f$) and their ratio $r= \rho_c/\rho_f$. 
\end{abstract}


\submitted{\SUST}

\maketitle

\section{Introduction}

To understand the physics of how magnetic flux enters a type-II superconductor under the influence of applied currents and fields is important not only for fundamental reasons but also for potential applications.  It is well known that magnetic flux enters in the form of quantized  vortices \cite{Abrikosov57}, each carrying magnetic flux $\phi_0 = h/2e$ concentrated within an area of the order of $\lambda^2$, where $\lambda$ is the London penetration depth.  In the case of greatest practical interest, when the intervortex spacing is much smaller than $\lambda$, the macroscopic magnetic flux density $\bm B$ (averaged over a few intervortex spacings) has magnitude $B = n \phi_0$, where $n$ is the local areal density of vortices.  

The orientation of $\bm B$ serves as a local reference direction.  The Lorentz force per unit volume on the vortex array is $\bm F = \bm J \times \bm B$.  When $F$ exceeds the local pinning force density $F_p$, the vortices generally move with a velocity $\bm v$ in the direction of $\bm F$, thereby producing an electric field \cite{Josephson65}  $\bm E = \bm B \times \bm v$.  In most cases of practical interest, such as in superconducting magnets, the magnetic induction in the superconductor produced by the self-fields is perpendicular to the  current direction.  To calculate quantities of practical interest, such as ac losses, one may then use the well-known critical-state theory \cite{Bean62,Bean64,Campbell72,Wilson83a,Carr83}, which assumes that $\bm B$ and $\bm J$ are perpendicular to each other and $\bm E$ is perpendicular to $\bm B$.  This critical-state theory characterizes the threshold for depinning in terms  of a critical current density $J_{c\perp} = F_p/B$, where here we use the subscript $\perp$ as a reminder that this refers to the case when $\bm J$ is perpendicular to $\bm B$. 

It has been known for many years, however, that when a type-II superconducting wire in a parallel magnetic field carries a current above the wire's critical current, a finite electric field appears along the length of the wire \cite{Walmsley72a}.  Inside the wire, there must be a place where $\bm J$, $\bm E$, and $\bm B$ are parallel, not perpendicular.  This parallel component of the electric field cannot be understood in terms of the collective motion of an array of unbroken vortices but can be understood as a consequence of flux-line cutting, the intersection and cross-joining of locally nonparallel vortices \cite{Campbell72,Walmsley72a}.  This is the analogue of the process of vortex-line reconnection in turbulent superfluid $^4$He, recently filmed by Paoletti {\it et al} \cite{Paoletti10}. Flux-line cutting can be triggered by the helical vortex expansion instability \cite{Clem77}, in which initially straight vortex lines subjected to a sufficiently large current density are susceptible to the growth of helical perturbations.  In this paper we use $J_{c\parallel}$ to denote the critical current at the threshold for flux-line cutting, where  we use the subscript $\parallel$ as a reminder that this refers to the case when  $\bm J$ is parallel to $\bm B$. 

Although the behaviour of type-II superconductors is well understood for the case that  $\bm J$ is perpendicular to $\bm B$, the theory for the case that $\bm J$ has components both perpendicular and parallel to $\bm B$ is still in a state of development.  Such a theory is needed, for example, to calculate the ac losses in power transmission cables consisting of helically wound layers of 2G coated-conductor composite tapes \cite{Clem10}.  If the tapes are tightly wrapped with no gaps between them, the self-field generated inside the cable has azimuthal and longitudinal components but no radial components.  In such a geometry,  the vortices in the tapes are subjected to a local current density with not only perpendicular components, which can drive the vortices either inward or outward in the radial direction, but also parallel components, which can lead to local helical instabilities and subsequent flux-line cutting \cite{Clem77}. It is the purpose of this paper to contribute towards the development of a critical-state theory accounting for both flux transport and flux cutting.

In Secs.\ \ref{GDCSM}, \ref{BMMsec}, \ref{OECSM}, \ref{EECSM}, and \ref{BLRMsec} we discuss the different predictions of five theoretical critical-state models for  how the  critical-current density $J_c$ and the electric field $\bm E$ (for $J$ just above $J_c$) should vary with the angle between  $\bm J$ and $\bm B$.  
While there has been previous experimental work on flux cutting in thin films with applied magnetic fields tilted out of plane but without in-plane angular rotation \cite{Blamire85a,Blamire85b}, we report in Sec.\ \ref{Expt} measurements with angular rotation of in-plane applied fields to test the theoretical predictions of the five models. 
We compare the experimental results with the predictions in Sec.\ \ref{Expt&Theory}, and we summarize our conclusions in Sec.\ \ref{Conclusions}.

\section{Generalized double-critical-state model (GDCSM) \label{GDCSM}}

In an attempt to describe the fundamentals of the behaviour of a type-II superconductor subjected to time-varying fields and currents  in situations where the current density $\bm J$ is not perpendicular to $\bm B$, a theory was proposed in  \cite{Clem82,Clem84,Perez85a,Perez85b,Clem86}.  This approach has been referred to  as the generalized double-critical-state model (GDCSM).  

Consider an infinite type-II superconducting slab centered on the $yz$ plane.   
Suppose that in the absence of an applied current, an applied magnetic field produces a magnetic flux density $\bm B_0 = B_0 \hat \alpha_0$ in the slab, where $\hat \alpha_0 =  \hat y \sin \alpha_0 + \hat z \cos \alpha_0$.  Consider now what happens when the slab carries an average current density $\bm {\overline J}$.    As a consequence of the self-field, vortices may be introduced, altering the magnetic flux density inside the slab to $\bm B(x) = B(x) \hat \alpha$, where $\hat \alpha(x)=  \hat y \sin \alpha(x) + \hat z \cos \alpha(x).$
The current density $\bm J$ also depends upon $x$.

\begin{figure}
\begin{center}
\includegraphics[width=5cm]{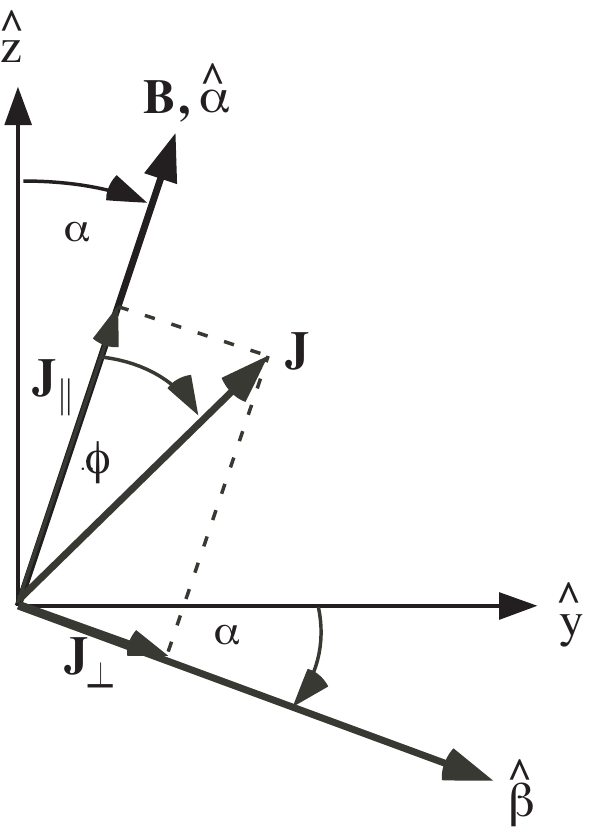}
\end{center}
\caption{Sketch of the unit vectors $\hat y$, $\hat z$, $\hat \alpha$, and $\hat \beta$, the magnetic induction $\bm B$, and the current density $\bm J$ and its components  $J_\parallel$ and $J_\perp$ along $\hat \alpha$ and $\hat \beta$.  $\phi$ is the angle between $\bm J$ and $\bm B$.  All vectors are parallel to the $yz$ plane. }
\label{fig1}
\end{figure}

Referring now to figure \ref{fig1}, let us resolve the current density and the electric field into components parallel and perpendicular to  $\bm B(x)$: $\bm J= J_y \hat y + J_z \hat z   = J_\parallel \hat \alpha + J_\perp \hat \beta$ and $\bm E = E_y \hat y + E_z \hat z = E_\parallel \hat \alpha + E_\perp \hat \beta$, where $\hat \beta = \hat y \cos \alpha -\hat z \sin \alpha = \hat \alpha \times \hat x$.  In high-$\kappa$ superconductors when $B_0$ is sufficiently above $B_{c1}$, it is a good approximation to take $B = \mu_0 H$ inside the superconductor.  In the steady state, Ampere's and Faraday's laws ($\bm J = \nabla \times \bm H,$ $\nabla \times \bm E = 0$, with all quantities independent of time) then yield \cite{Clem86}
\begin{eqnarray}
J_\parallel &=& J \cos \phi = \frac{B}{\mu_0}\frac{d \alpha}{d x},\\
J_\perp &=&J \sin \phi = -\frac{1}{\mu_0}\frac{d B}{d x},
\end{eqnarray}
\begin{eqnarray}
\frac{d E_y}{dx} &=&0, \\
\frac{d E_z}{dx} &=&0,
\end{eqnarray}
\begin{eqnarray}
E_\parallel \frac{d \alpha}{d x}+\frac{d E_\perp}{d x} &=& 0,\\
E_\perp \frac{d \alpha}{d x}-\frac{d E_\parallel}{d x} &=& 0.
\end{eqnarray}

The GDCSM was based on the fundamental ideas that a nonvanishing $E_\perp$ can occur only when the magnitude of $J_\perp$ is above $J_{c\perp}$, the critical current density for flux depinning, and that a nonvanishing $E_\parallel$ can occur only when the magnitude of $J_\parallel$ is above $J_{c\parallel},$ the critical current density for flux cutting.  It was assumed  in   \cite{Clem86} that  the superconducting material was isotropic and that the pinning centers were isotropically distributed, such that $J_{c\perp}$ and $J_{c\parallel}$ could depend upon the  temperature $T$ but only upon the magnitude $B$ of the magnetic induction $\bm B$ inside the sample.  For thin samples, the current-induced self-field  is much smaller than the applied magnetic induction $\bm B_0$, so that $J_{c\perp}$ and $J_{c\parallel}$ were taken as constants independent of  $x$.  The possible complications of surface pinning or surface barriers were ignored.
One of the most important assumptions of the GDCSM was that flux depinning and flux cutting do not affect each other; i.e., it was implicitly assumed that $J_{c\perp}$ was independent of $J_{\parallel}$ and $J_{c\parallel}$ was independent of $J_{\perp}$. 

\begin{figure}
\begin{center}
\includegraphics[width=8cm]{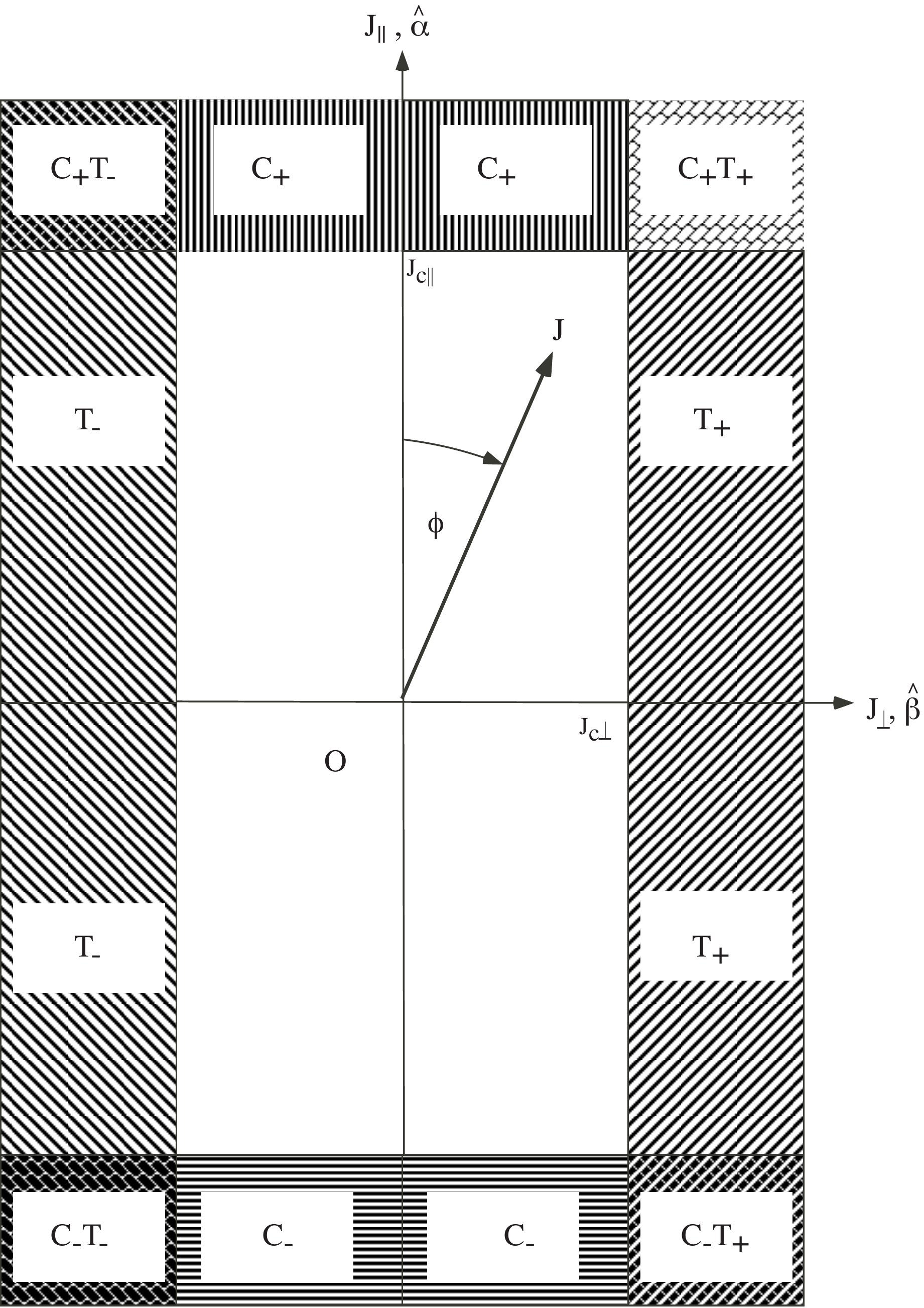}
\end{center}
\caption{Behaviour of a vortex array of flux density $\bm B$ (which is parallel to the $J_{\parallel}$ axis) as a function
of the magnitude $J$ of the current density $\bm J$ and the angle $\phi$ of $\bm J$ relative to $\bm B$
according to the generalized double-critical-state
model \cite{Clem82,Clem84,Perez85a,Perez85b} (GDCSM), in which
$J_{c\perp}$ and $J_{c\parallel}$ are both assumed to be constants. }
\label{fig2}
\end{figure}

Figure \ref{fig2} provides a graphical display of the behaviour of the
vortex array as a function of $J_{\perp} = J \sin \phi$ and
$J_{\parallel} = J
\cos \phi$ according to the GDCSM.  In the middle of the open rectangle, the 0 zone, for which $|J_{\perp}| < J_{c\perp}$ and
$|J_{\parallel}| <  J_{c\parallel}$, neither flux transport (depinning)
nor flux-line cutting occurs; i.e., both $E_{\perp}$ and
$E_{\parallel}$ are zero.   Flux transport (without flux-line cutting) occurs only in the zones labelled by T$_+$ ($J_{\perp} > J_{c\perp}$ and
$E_{\perp}>0$) or T$_-$ ($J_{\perp} <- J_{c\perp}$ and
$E_{\perp}<0$).  Flux-line
cutting  (without flux transport) occurs only in zones labelled by C$_+$ ($J_{\parallel} > J_{c\parallel}$ and
$E_{\parallel} >  0$) or  C$_-$ ($J_{\parallel} <- J_{c\parallel}$ and $E_{\parallel} < 0$).  Simultaneous flux transport ($|J_{\perp}| > J_{c\perp}$ and
$|E_{\perp}|>0$) and flux-line cutting ($|J_{\parallel}| > J_{c\parallel}$ and $|E_{\parallel}| > 0$) occurs  only in the zones labelled by C$_+$T$_+$,  C$_-$T$_+$,  C$_-$T$_-$, and  C$_+$T$_-$.

\begin{figure}
\begin{center}
\includegraphics[width=8cm]{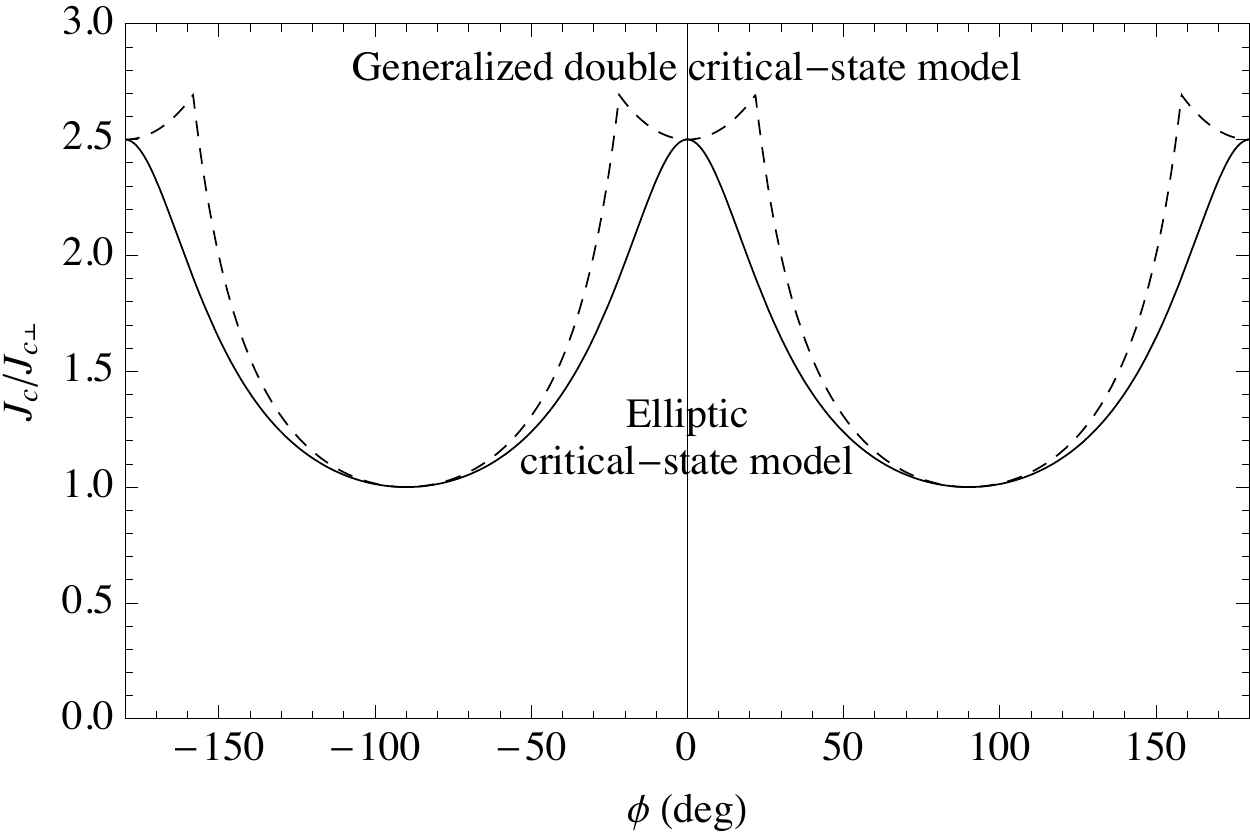}
\end{center}
\caption{Plot of $J_c$ (in units of $J_{c\perp}$) vs
$\phi$ for two theoretical models accounting for flux-line cutting when
$J_{c\parallel}/J_{c\perp}= 2.5 $, for which 
$\phi_c = \tan^{-1}(0.4) = 0.381 \;{\rm rad} = 21.8^\circ$.  The dashed curve shows $J_c/J_{c\perp}$ according to the generalized double critical-state model (GDCSM) \cite{Clem86},
 (\ref{Edepin}) and (\ref{Ecut}), and the solid curve shows 
$J_c/J_{c\perp}$ according to
the elliptic critical-state model \cite{Romero03a,Romero03b,Romero04}, (\ref{Jcellipse1}) and (\ref{Jcellipse}). }
\label{fig3}
\end{figure}

According to the GDCSM, for given values of $J_{c\perp}$ and $J_{c\parallel}$, the critical
current density $J_c$ at the first onset of a nonvanishing electric field
depends upon the angle $\phi$.  For $\phi = \pi/2$ ($\bm J \perp \bm B$), the critical
current density for the onset of an electric field is the usual depinning
critical current density, such that $J_c = J_{c\perp}$.  On the other
hand, for $\phi = 0$ ($\bm J \parallel \bm B$), the critical current density for the onset of an
electric field is the current density at the threshold for  flux-line
cutting, and $J_c = J_{c\parallel}$.  For other values of $\phi$, we
have 
\begin{eqnarray} 
J_c &=& \frac{J_{c\perp}}{|\sin\phi|}, \;\; |\tan\phi|
\ge \tan\phi_c, \label{Edepin} \\
&=&\frac{J_{c\parallel}}{|\cos\phi|},
\;\; |\tan\phi| \le \tan\phi_c,
\label{Ecut}
\end{eqnarray} 
where $\tan\phi_c = J_{c\perp}/J_{c\parallel}$.  As is evident from figure 2 and 
(\ref{Edepin}) and (\ref{Ecut}), this model predicted that
$J_c$ as a function of
$\phi$ has cusplike maxima at
$\phi =
\pm \phi_c$ and $\phi = \pm (\pi - \phi_c)$, where $J_{cmax} =
\sqrt{J_{c\perp}^2+J_{c\parallel}^2}$.  For example, the dashed curve in figure \ref{fig3}   maps the magnitude of
$\bm J$ as a function of $\phi$ as the tip of the vector $\bm J$ traces
the boundary of the 0 zone shown in figure \ref{fig2}.  The rectangular shape of
the 0 zone in figure \ref{fig2} arises from the assumptions of the
GDCSM \cite{Clem82,Clem84,Perez85a,Perez85b,Clem86} that (a) the threshold
$J_{c\perp}$ for depinning is independent of the current density 
$J_{\parallel}$ parallel to the vortices and (b) the threshold
$J_{c\parallel}$ for flux-line cutting is independent of the current
density 
$J_{\perp}$ perpendicular to the vortices. 

\begin{figure}
\begin{center}
\includegraphics[width=8cm]{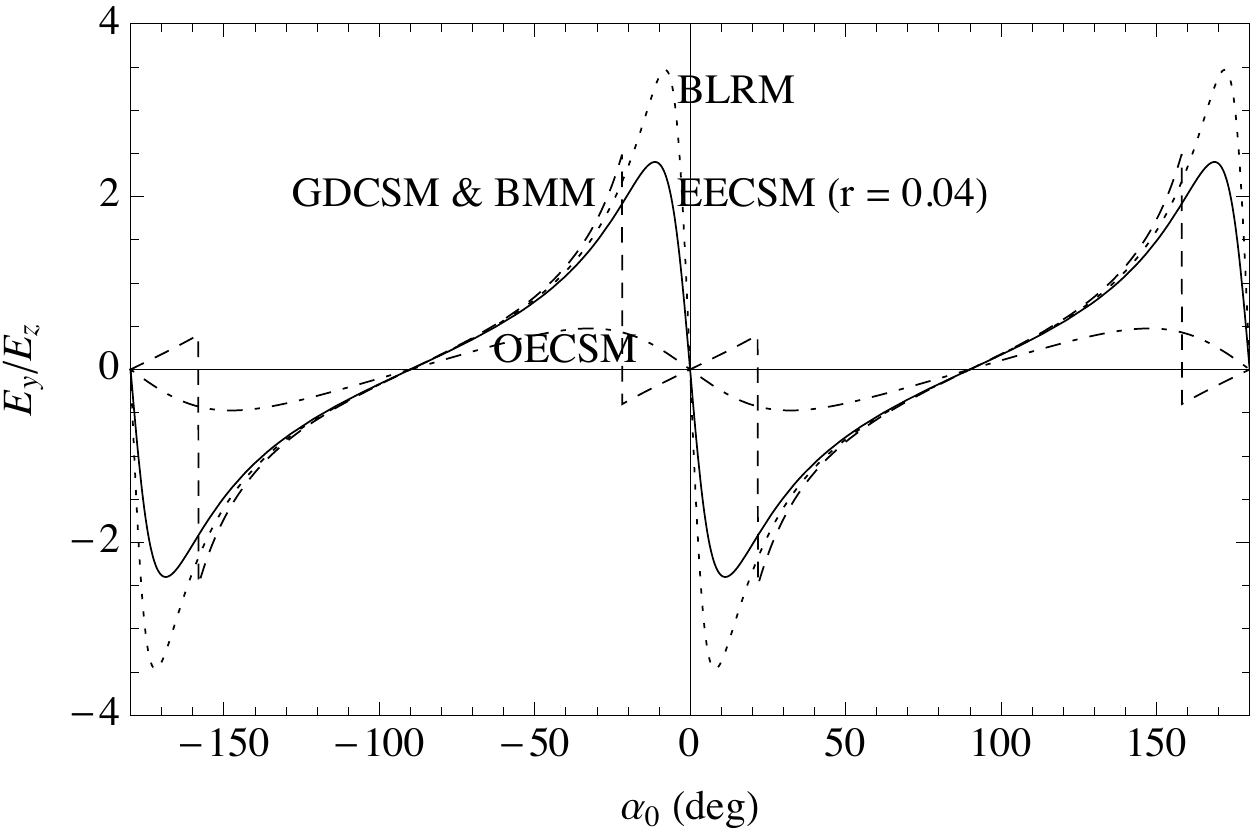}
\end{center}
\caption{
Plots of the electric field ratio $E_y/E_z$ vs
$\alpha_0$  for five theoretical models accounting for flux-line cutting.  The predictions of the generalized double critical-state model and the Brandt-Mikitik model [GDCSM \& BMM, dashed,  (\ref{EybyEzTransport}) and (\ref{EybyEzCutting})] and the original elliptic critical-state model [OECSM, dot-dashed,  (\ref{tanpsiE})]  are all shown for
$\tan\phi_c=J_{c\perp}/J_{c\parallel}=0.4$, for which 
$\phi_c = \tan^{-1}(J_{c\perp}/J_{c\parallel})  = 21.8^\circ$.  Also shown is the  behaviour of $E_y/E_z$ predicted by the extended elliptic critical-state model [EECSM, solid,   (\ref{tanpsiEcreepa})] for an assumed value of $r = 0.04$, and that of the BLRM [dotted, (\ref{BLRmodel})] for $\tan^2\phi_c=(J_{c\perp}/J_{c\parallel})^2=0.02$.}
\label{fig4}
\end{figure}

Another prediction of the GDCSM \cite{Clem86} was that, in accord with figure \ref{fig2}, measurements of the angle of the electric field $\bm E$ just above the critical current as a function of the angle $\phi$ between $\bm J$ and $\bm B_0$ would show sharp changes  at  $|\phi| =
\phi_c = \tan^{-1}(J_{c\perp}/J_{c\parallel})$ and $|\phi| = \pi - \phi_c$.  If $\bm J$ is constrained by geometry to flow only in the $z$ direction, then just above the critical current density, where $\bm J = J_c \hat z$ and $\bm B = B_0(\hat y \sin \alpha_0 + \hat z \cos \alpha_0)$, we have $\phi = - \alpha_0$, so that $\bm E = E_y \hat y +E_z \hat z = E_\parallel \hat \alpha + E_\perp \hat \beta$, where $\hat \alpha =\hat y \sin \alpha_0 + \hat z \cos \alpha_0$ and $\hat \beta = \hat y \cos \alpha_0 - \hat z \sin \alpha_0$.  The GDCSM prediction was that when $\phi_c < \alpha_0 < \pi - \phi_c$ or  $-\pi+\phi_c < \alpha_0 < - \phi_c$, the behaviour for $J > J_c$ would be dominated by flux transport, such that $E_\parallel = 0$, $\bm E \perp \bm B_0$, $\bm E  = E_\perp \hat \beta = E_\perp (\hat y \cos \alpha_0 - \hat z \sin\alpha_0)$, and 
\begin{equation}
E_y/E_z = -\cot \alpha_0.  
\label{EybyEzTransport}
\end{equation}
On the other hand,  when $-\phi_c < \alpha_0 <\phi_c$, the GDCSM prediction was that the behaviour  for $J > J_c$ would be dominated by flux cutting, such that $E_\perp = 0$, $\bm E \parallel \bm B_0$, $\bm E  = E_\parallel \hat \alpha = E_\parallel (\hat y \sin \alpha_0 + \hat z \cos\alpha_0)$, and 
\begin{equation}
E_y/E_z = \tan \alpha_0.  
\label{EybyEzCutting}
\end{equation} 
The dashed curve in figure \ref{fig4} shows the behaviour of $E_y/E_z$ predicted by the GDCSM.

\section{Brandt and Mikitik model (BMM), extending the GDCSM model  \label{BMMsec}}

Brandt and Mikitik \cite{Brandt07} proposed an extension (here called the BMM) of the GDCSM, in which the threshold for flux-line cutting is reduced as a function of the magnitude of $J_\perp$ and the threshold for depinning is reduced as a function of the  magnitude of $J_\parallel$.  If these thresholds were bent enough, the shape of the surface of $J_c(\phi)$ could be made to resemble an ellipse, as suggested by figure 1a of \cite{Brandt07}, thereby eliminating the cusplike behavior predicted by the GDCSM. 
In fact, if the BMM $J_c(\phi)$ curve had the shape of an ellipse, it would be in exact agreement with the solid curve shown in figure \ref{fig3}.

 However, the BMM retained the assumption of the GDCSM that    the electric field  $\bm E$ is parallel to the $J_\parallel$ axis on the flux-cutting-threshold portions of the $J_c(\phi)$ curve closest to the $J_\parallel$ axis and that $\bm E$ is parallel to the $J_\perp$ axis on the flux-transport-threshold portions of the $J_c(\phi)$ curve closest to the $J_\perp$ axis.  Thus, in common with the GDCSM, the BMM predicts a 90 degree change in the direction of $\bm E$ when $\phi$ passes through the point of intersection of the curves representing the thresholds for flux cutting and flux transport.  If this intersection occurs when $|J_\perp/J_\parallel| = J_{c\perp}/J_{c\parallel}$, then the shape of the curve of $E_y/E_z$ curve agrees exactly with that of the GDCSM, as shown in figure \ref{fig4}.

\section{Original elliptic critical-state model (OECSM) \label{OECSM}}

Experimental results of Fisher {\it et
al} \cite{Fisher97,Fisher00} on the time evolution of the static magnetic moment of a superconducting plate subjected to an alternating magnetic field applied
perpendicular to a dc magnetic field  were found not to be well described
theoretically by calculations based on the GDCSM.  
However,
Romero-Salazar and P\'erez-Rodr\'iguez \cite{Romero03a,Romero03b,Romero04}
introduced an elliptic critical-state model, which they  found
yielded good theoretical agreement with the experiments of 
\cite{Fisher97,Fisher00}.

\begin{figure}
\begin{center}
\includegraphics[width=10
cm]{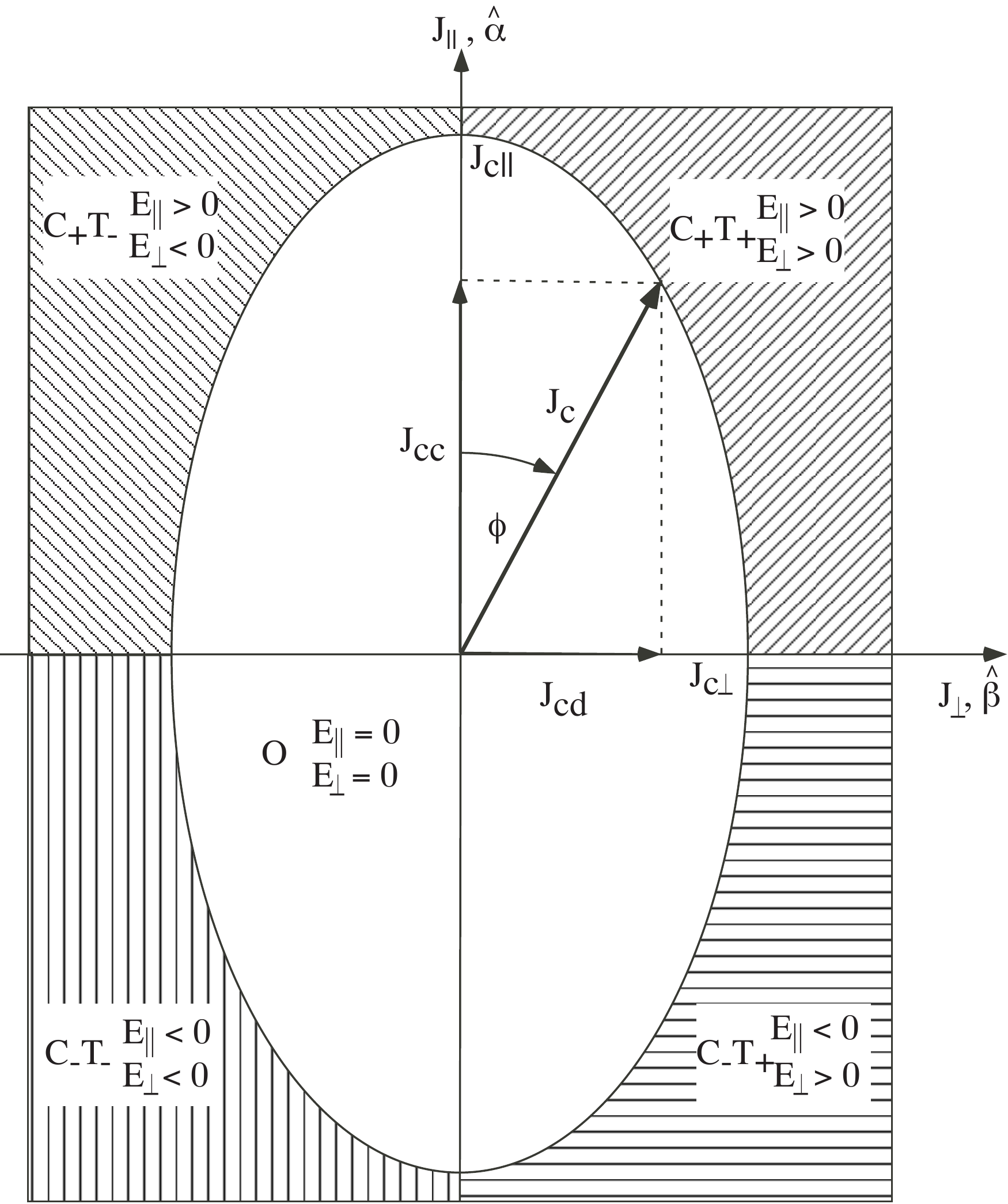}
\end{center}
\caption{Behaviour of a vortex array of flux density $\bm B$ as a function
of the magnitude $J$ and angle $\phi$ of the current density $\bm J$ relative to the direction of $\bm B$ when
flux-line cutting and depinning interact and the critical current density
is given by the elliptic critical-state model of 
\cite{Romero03a,Romero03b,Romero04}.  Here $\bm J$ is shown at the critical current density, where $J = J_c,$  $J_{cc}=J_c\cos\phi$ is the threshold for flux cutting, and $J_{cd}=J_c\sin\phi$ is the threshold for flux transport (depinning).}
\label{fig5}
\end{figure}

Figure \ref{fig5} provides a graphical display of the behaviour of the vortex array
as a function of $J_{\perp} = J \sin \phi$ and $J_{\parallel} = J
\cos \phi$ according to the elliptic critical-state
model \cite{Romero03a,Romero03b,Romero04}. 
We interpret this behaviour as follows:  In the 0 zone inside the ellipse described by 
\begin{equation}
\frac{\sin^2\phi}{J_{c\perp}(B)^2}+
\frac{\cos^2\phi}{J_{c\parallel}(B)^2}=\frac{1}{J_c(B,\phi)^2}
\label{Jcellipse1}
\end{equation}
or 
\begin{equation}
J_c(B,\phi)=1/\sqrt{\frac{\sin^2\phi}{J_{c\perp}(B)^2}+
\frac{\cos^2\phi}{J_{c\parallel} (B)^2}},
\label{Jcellipse}
\end{equation}
neither flux transport (depinning) nor flux cutting occurs ($E_{\perp} = 0$ and
$E_{\parallel}=0$).
Flux transport, for which the vortices are
depinned ($E_{\perp}\ne 0$), occurs everywhere outside the ellipse (except when
$J_{\perp}=0$) in zones with labels including the symbol T$_+$ ($E_\perp > 0$) or T$_-$ ($E_\perp < 0$).  Flux-line cutting, for which
$E_{\parallel} \ne  0$, occurs everywhere outside the ellipse (except when
$J_{\parallel}=0$) in zones with labels including the symbol C$_+$  ($E_{\parallel} > 0$) or 
C$_-$ ($E_{\parallel} < 0$).  In this model, both flux transport and flux-line cutting occur simultaneously nearly everywhere outside the ellipse of $J_c(\phi)$ vs $\phi$.  The solid curve in figure \ref{fig3} gives an
example of the behaviour of $J_c$ vs $\phi$ using the elliptic
critical-state model.

We note here that the behaviour at the critical current according to the elliptic critical-state model (see figure \ref{fig5}) requires that flux depinning and flux cutting be interdependent.   
To formulate this, we introduce the following new notation for the critical current densities at the thresholds for flux depinning $J_{cd}(B,\phi)$ and flux cutting $J_{cc}(B,\phi)$, where $\phi$ is the angle between $\bm J$ and $\bm B$, as in figure \ref{fig1}.  
Our interpretation of the underlying physics of the elliptic critical-state
model \cite{Romero03a,Romero03b,Romero04} is that  (a) the threshold
$J_{cd}$ for depinning is a monotonically decreasing function of the magnitude of 
$J_{\parallel}$, decreasing from $J_{c\perp}$ when $J_\parallel = 0$ to zero when $|J_\parallel| = J_{c\parallel}$, 
and (b) the threshold
$J_{cc}$ for flux-line cutting is a monotonically decreasing function of the magnitude of
$J_{\perp}$,   decreasing from $J_{c\parallel}$ when $J_\perp = 0$ to zero when $|J_\perp| = J_{c\perp}$.   
 In summary, at the critical current density, $J = J_c(B,\phi)$, and $\bm J$
lies on the ellipse of   (\ref{Jcellipse}) as shown in figure \ref{fig5}, such that $|J_{\perp}| = J_{cd}(B,T) = J_c(B,\phi)
|\sin
\phi|$,  $|J_{\parallel}| =J_{cc}(B,T)  = J_c(B,\phi)
|\cos \phi|$, and 
\begin{equation} 
J_c(B,\phi) = \frac{J_{c\perp}(B)}{\sqrt{\sin^2 \phi+\tan^2
\phi_c \cos^2
\phi}},
\label{Jcelliptic}
\end{equation} 
where $\tan \phi_c = J_{c\perp}(B)/J_{c\parallel}(B)$.
Note that within this model $J_{c\perp}(B) = J_c(B,\pi/2)$ and $J_{c\parallel}(B) = J_c(B,0)$.

Further  fundamental assumptions of the original elliptic critical-state model \cite{Romero03a,Romero03b,Romero04} are that the electric field obeys  
\begin{eqnarray}
E(J) &=&0, \; \;\;\;\;\;\;\;\;\;\;\;\;\;\;\;\;\;\;\;\;\;\;J\le J_c(B,\phi),\\
&=&\rho[J-J_c(B,\phi)],\;J> J_c(B,\phi),
\end{eqnarray}
and that the components of $\bm E$ perpendicular and parallel to $\bm B$ obey \cite{Romero03a,Romero03b,Romero04}
\begin{eqnarray}
E_\perp/E & = &J_\perp/J_{c\perp}, \label{Eperpelliptic} \\
E_\parallel/E & = &J_\parallel/J_{c\parallel}. \label{Eparallelelliptic}
\end{eqnarray}
If the current is constrained by geometry to flow only in the $z$ direction, then at the critical current density, when  $\bm J = \hat z J_c$ and $\bm B = B_0 (\hat y \sin \alpha_0 +\hat z \cos \alpha_0)$, we have $\phi = -\alpha_0$, such that  $J_\parallel = J_c \cos \alpha_0$ and $J_\perp = - J_c \sin \alpha_0$ (see figure \ref{fig1}) and
\begin{eqnarray}
E_\perp/E & = &-(J_c/J_{c\perp})\sin \alpha_0,
\label {EperpJc}\\
E_\parallel/E & = &(J_c/J_{c\parallel})\cos \alpha_0.
\label{EparallelJc}
\end{eqnarray}
Since $\bm E = E_y \hat y + E_z \hat z = E_\parallel \hat \alpha + E_\perp \hat \beta$ and $E_\parallel/E_\perp = -(J_{c\perp}/J_{c\parallel})/\tan\alpha_0$,  we have 
\begin{equation}
\frac{E_y}{E_z}= \frac{(E_\parallel/E_\perp)\tan \alpha_0+1}{(E_\parallel/E_\perp)-\tan\alpha_0}= \frac{(\tan\phi_c-1)\tan \alpha_0}{\tan\phi_c+\tan^2\alpha_0},
\label{tanpsiE}
\end{equation}
where $\tan\phi_c=J_{c\perp}/J_{c\parallel}$.  Experiments generally yield $\tan\phi_c < 1$.  
According to this form of the elliptic critical-state model,  as a function of $\tan \alpha_0$, $E_y/E_z$ has extrema (a maximum and a minimum) when   $\tan \alpha_0 = \mp \sqrt{\tan\phi_c}$, where $E_y/E_z = \pm |1-\tan\phi_c|/2\sqrt{\tan\phi_c}$.
The dot-dashed curve in figure \ref{fig4} shows the behaviour of $E_y/E_z$ predicted by the original elliptic critical-state model \cite{Romero03a,Romero03b,Romero04},  (\ref{tanpsiE}).

\section{Extended elliptic critical-state model (EECSM)\label{EECSM}}

However, the original elliptic critical-state model  \cite{Romero03a,Romero03b,Romero04} suffers from an important deficiency.  By summing the squares of  (\ref{Eperpelliptic}) and (\ref{Eparallelelliptic}), we obtain 
\begin{equation}
(\frac{E_\perp}{E})^2 + (\frac{E_\parallel}{E})^2 = 1 = 
(\frac{J_\perp}{J_{c\perp}})^2 + (\frac{J_\parallel}{J_{c\parallel}})^2.
\end{equation}
The right-hand side can be equal to one only when $J = J_c$, as can be seen by making use of  (\ref{EperpJc}), (\ref{EparallelJc}), and (\ref{Jcellipse1}). However, when $J = J_c$, we must have $E = 0$. Therefore, additional equations are required in order to apply the elliptic critical-state model to finite values of the electric field when $J > J_c$, where the dynamical processes of flux transport and flux cutting become important.

We therefore propose an extension of the elliptic critical-state model to account for dissipative processes when $J > J_c$  by writing
\begin{eqnarray}
E_\perp & = &\rho_f J_\perp,
\label {Eflow}\\
E_\parallel & = &\rho_c J_\parallel,
\label{Ecut2}
\end{eqnarray}
where $\rho_f$ is a  nonlinear function of $J$, $B$, $\alpha_0$, and $T$ with a positive value describing flux flow and $\rho_c$ is another  nonlinear function of $J$, $B$, $\alpha_0$, and $T$ with a positive value describing flux cutting.  In the absence of flux creep, $\rho_f$ is zero when $|J_\perp|$ is less than the threshold $J_{cd}$ for depinning, and  $\rho_c$ is zero when $|J_\parallel|$ is less than the threshold $J_{cc}$ for flux cutting.  To clarify what we mean by a threshold for flux-line cutting, consider starting with the sample carrying a current for which $J_\parallel = 0$ and $0 < J_\perp < J_{c\perp}$, such that $\bm E = 0$.  If $J_\parallel$ is increased above the threshold value $J_{cc}$, flux-line cutting begins ($E_\parallel > 0$), but the  elliptic critical-state model (and the experiments to be presented later) show that this is accompanied by flux transport ($E_\perp > 0$).  In other words, when $J > J_c(\phi)$,  flux cutting and flux transport occur simultaneously; the thresholds for flux cutting and flux transport are intimately linked.

If the current is constrained by geometry to flow only in the $z$ direction, then at the critical current density, when  $\bm J = \hat z J$ and $\bm B = B_0 (\hat y \sin \alpha_0 +\hat z \cos \alpha_0)$, we have $\phi = -\alpha_0$, such that  $J_\parallel = J \cos \alpha_0$ and $J_\perp = - J \sin \alpha_0$ (see figure \ref{fig1}). 
Since $\bm E = E_y \hat y + E_z \hat z = E_\parallel \hat \alpha + E_\perp \hat \beta$, where $\hat \alpha= \hat y \sin \alpha_0 + \hat z \cos \alpha_0$ and $\hat \beta = \hat y \cos \alpha_0 - \hat z \sin \alpha_0$,  and $E_\parallel/E_\perp = -(\rho_c/\rho_f)/\tan\alpha_0$, we obtain, making use of  (\ref{Eflow}) and (\ref{Ecut2}),
\begin{equation}
\frac{E_y}{E_z} = \frac{(E_\parallel/E_\perp)\tan \alpha_0+1}{(E_\parallel/E_\perp)-\tan\alpha_0}=\frac{(r-1)\tan \alpha_0}{r+\tan^2\alpha_0},
\label{tanpsiEcreepa}
\end{equation}
where now $r = \rho_c/\rho_f$, which  must be a positive number.

Note that   (\ref{tanpsiEcreepa}) has the same mathematical form as   (\ref{tanpsiE}), except that $\tan \phi_c$ is replaced by $r$.  According to this extension of the theory, $E_y/E_z$ does not depend at all upon $\tan \phi_c = J_{c\perp}/J_{c\parallel}$ but instead depends upon $r$, the ratio of the effective flux-cutting resistivity to the effective flux-flow resistivity.  If $r$ is independent of $\tan \alpha_0$, $E_y/E_z$ has extrema (a maximum and a minimum) when   $\tan \alpha_0 = \mp \sqrt{r}$, where $E_y/E_z = \pm |1-r|/2\sqrt{r}$.  The parameter $r$ thus plays a crucial role in describing the dynamical properties of the mixed state above the critical current density $J_c$.  Unfortunately, experiments to date typically do not report values of $\rho_c$ and $\rho_f$, from which $r = \rho_c/\rho_f$ could be independently measured.

The solid curve in figure \ref{fig4} shows the behaviour of $E_y/E_z$ predicted by this extended elliptic critical-state model   (\ref{tanpsiEcreepa}) for an assumed value of $r = 0.04$.

\section{Bad\'ia-Maj\'os, L\'opez, and Ruiz model (BLRM) \label{BLRMsec}}

In \cite{Badia09}, Bad\'ia-Maj\'os {\it et al} argued that for $J$ just above $J_c(\phi)$, the direction of the electric field should be perpendicular to the $J_c(\phi)$ curve. 
Under the same conditions leading to Eqs.\ (\ref{tanpsiE}) and (\ref{tanpsiEcreepa}), their model (here called the BLRM) leads to the expression
\begin{equation}
\frac{E_y}{E_z} = \frac{1}{J_c(\alpha_0)}\frac{dJ_c(\alpha_0)}{d\alpha_0}=\frac{(\tan^2\phi_c-1)\tan \alpha_0}{\tan^2\phi_c+\tan^2\alpha_0},
\label{BLRmodel}
\end{equation}
where the latter equation applies when $J_c(\phi)$ is given by the ellipse of (\ref{Jcellipse}) (see  figure \ref{fig5}) and $\tan\phi_c=J_{c\perp}/J_{c\parallel}$. The dotted curve in figure \ref{fig4} shows the behaviour of $E_y/E_z$ predicted by the BLRM  for an assumed value of $\tan^2\phi_c=(J_{c\perp}/J_{c\parallel})^2 = 0.02.$

\section{Experiment \label{Expt}}

We next report experiments to test the fundamental assumptions of the above-described theories.  While the predictions have been made for an infinite superconducting slab, we have carried out experiments on a current-carrying superconducting film of finite dimensions subjected to an in-plane magnetic field.  
We have assumed that the effects of finite dimensions and self-fields are small enough that the theoretically predicted effects dominate and the experiments can determine which of the proposed theories is correct.

\begin{figure}
\begin{center}
\includegraphics[width=4cm]{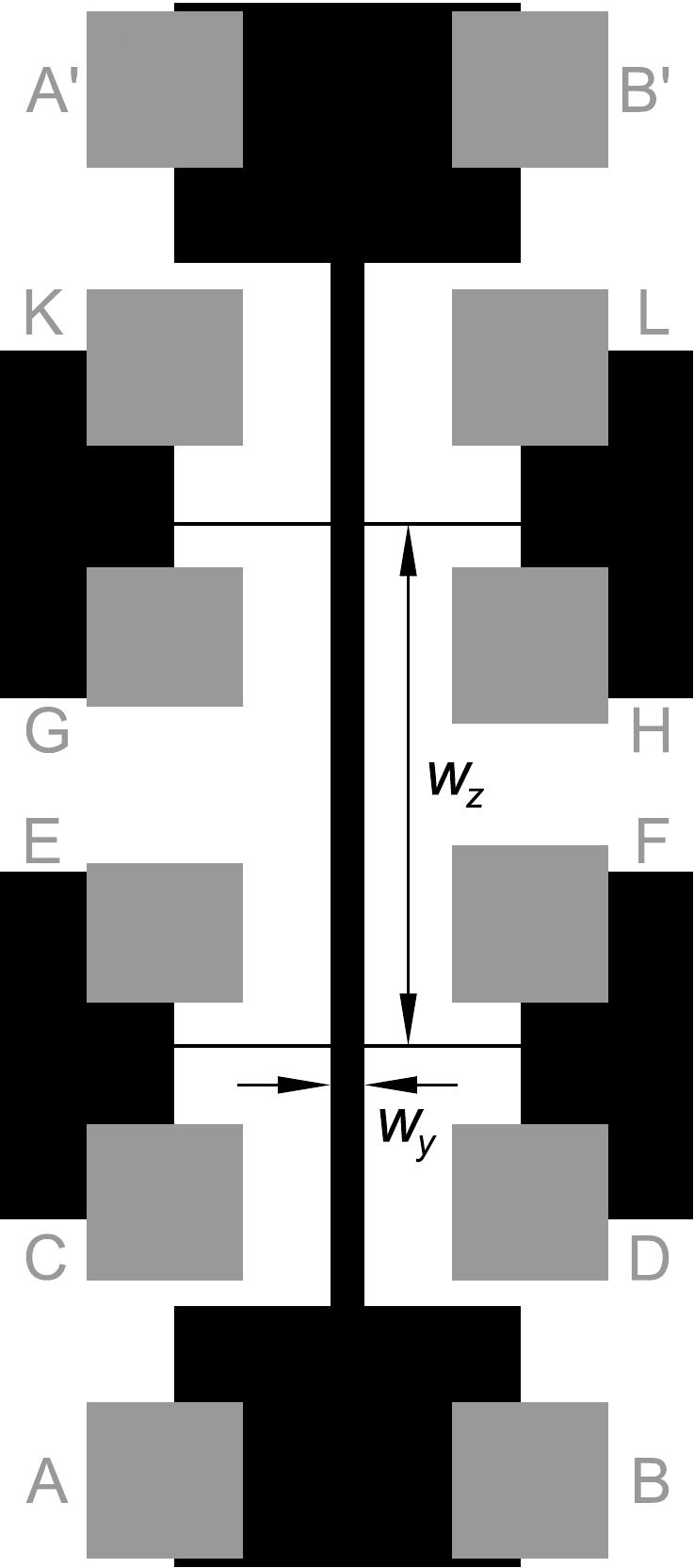}
\end{center}
\caption{The pattern used. The track is shown in black and the voltage/current taps in gray with their labels A, B, C, D, E, F, G, H, K, L, A', and B'. Voltage measurements were made over the length $w_z = 3000\um$ and width $w_y = 200\um$.}
\label{fig:sample_pattern}
\end{figure}

A sketch of the measured device, a YBCO film of thickness $500\nm$, is shown in figure \ref{fig:sample_pattern}. The track is of width   $w_y = 200\um$, and two sets of voltage taps are positioned along its length, $w_z = 3000\um$ apart from each other. The voltage/current taps are labelled A, B, C, D, E, F, G, H, K, L, A', B'.  
A two-axis goniometer \cite{Herzog94} was employed to measure the longitudinal and transverse voltage simultaneously by a four-terminal technique, using the two channels of the employed voltmeter, while a magnetic field was swept in-plane.
For all measurements the current $I$ was injected through the taps A and B and taken out through A' and B', such that the current direction was (A, B) $\rightarrow$ (A', B'), i.e., in the positive $z$ direction. The current was applied in short pulses to minimize sample heating.  Longitudinal and transverse voltages were measured using combinations of the taps C, \dots, L.

The applied field was oriented to be in the plane of the film, such that angle $\theta$ between the applied field and the film normal $\vec{n}$ was as close as possible to 90$\degree$, as described below.  The current density $\bm J$ was constrained to flow in the $z$ direction, while the angle $\alpha = \alpha_0$ of the applied magnetic induction $\bm B_0$ was oriented as shown in figure \ref{fig1}, such that the angle of $\bm J$ relative to $\bm B_0$ was $\phi = -\alpha_0$.

In order to align the sample such that $\bm B_0||ab$, measurements of $J_c(\theta)$ were performed at $\phi=\pm 90\degree$, corresponding to maximum Lorentz force ($\bm B_0 \perp \bm J$). 
The angle $\theta = 90\degree$ (where $\bm B_0||ab$) was assumed to be reached where $J_c(\theta)$ reached a maximum, and the following in-plane scans were performed with this value of $\theta$.

For the reported measurements, $J_c$ was determined from $V_z(I)$ curves only, where $V_z$ is the voltage measured along the length of the track (e.g.,\ by using taps C and G). For each value of $\alpha_0$, $I_c$ was  determined as the current where $V_z$ reaches the selected  voltage criterion $V_c$. The transverse voltage $V_y(I)$ was measured simultaneously (e.g., using taps G and H).  The voltages were then  converted to electric fields [$E_z = V_c / w_z$ and $E_y = V_y(I_c) / w_y$] in order to obtain $E_y/E_z$ as a function of $\alpha_0$.

\begin{figure}
\begin{center}
\includegraphics[width=8cm]{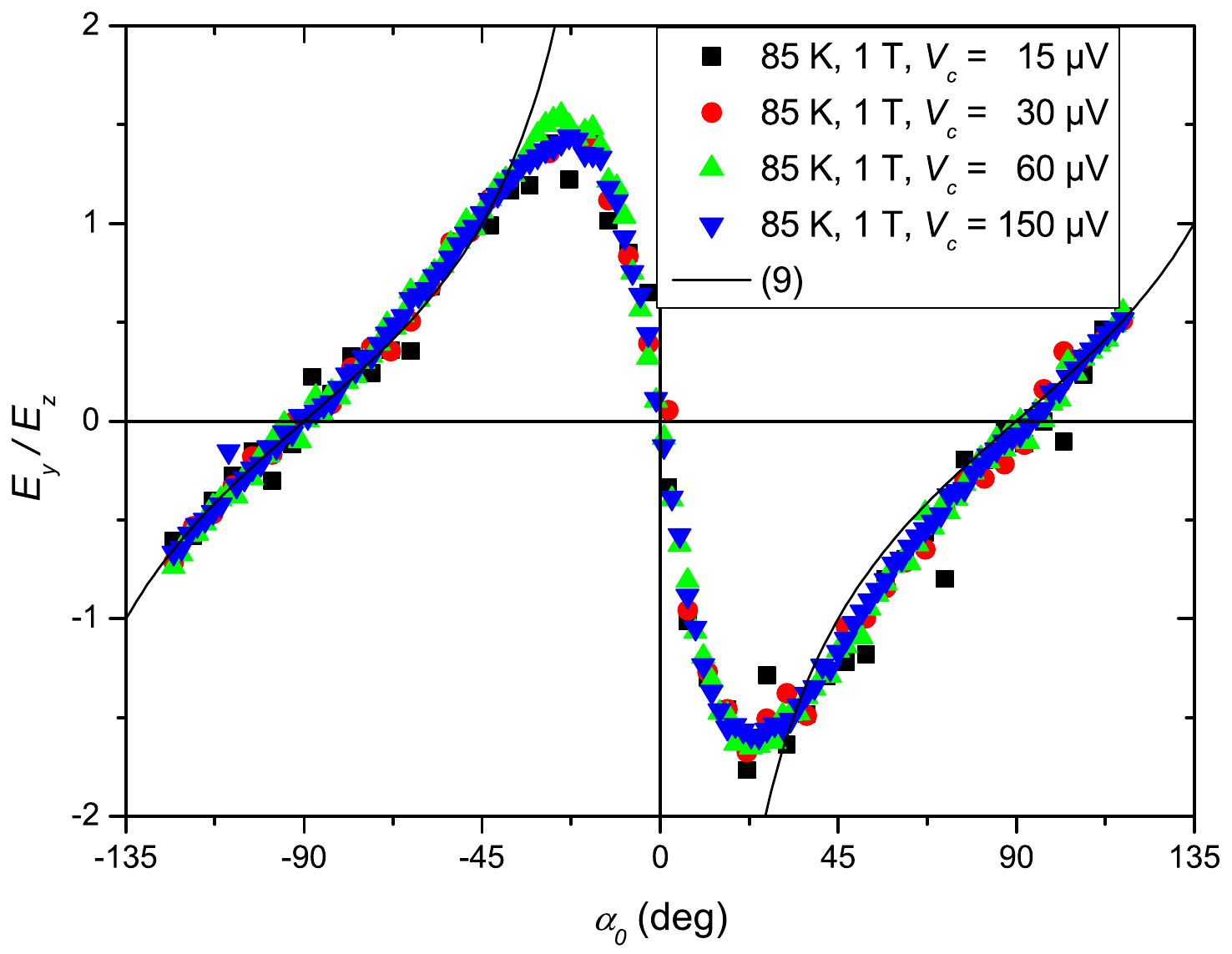}
\end{center}
\caption{The measured ratio $E_y/E_z$ vs $\alpha_0$ at $T = 85\K$ and $\mu_0H = 1\T$.  The solid curve is  $E_y/E_z = -\cot\alpha_0$  (\ref{EybyEzTransport}), expected due to flux transport in the absence of flux-line cutting.}
\label{fig:Ez_CGHG_85K_1T}
\end{figure}

Figure~\ref{fig:Ez_CGHG_85K_1T} presents plots of $E_y/E_z$ vs $\alpha_0$ obtained at $T = 85\K$ and $\mu_0H = 1\T$ for voltage criteria $V_c = 15\uV$ (square symbols), $30\uV$ (dots), $60\uV$ (triangles),  and $150\uV$ (inverted triangles).  Note that  the scatter in the values of $E_y/E_z$ becomes smaller  with increasing values of $V_c$.   In the raw data, $E_y/E_z$ reached zero at a value of $\alpha_0 = +3\degree$. This could be understood by  the fact that the alignment of the in-plane angle was performed by visual inspection prior to sample mounting. Consequently, we shifted the curves  by $-3\degree$ in $\alpha_0$, so that $E_y/E_z = 0$ at $\alpha_0 = 0\degree$ in all the plots shown here and later.  The solid curve in  figure~\ref{fig:Ez_CGHG_85K_1T} shows the function $E_y/E_z = -\cot\alpha_0$  (\ref{EybyEzTransport}), the result that would be expected for no flux-line cutting; i.e.,\ when $E_\parallel = 0$, $\bm E \perp \bm B_0$, and $\bm E = E_y \hat y + E_z \hat z = E_\perp \hat \beta=E_\perp ( \hat y \cos \alpha_0 -\hat z \sin \alpha_0)$. (Note from  figure \ref{fig1} that in this experiment  $\phi=-\alpha_0$, so that when $\alpha_0 > 0$, we have $E_z > 0$,  $E_y <0$, and $E_\perp < 0$.)

\begin{figure}
\begin{center}
\includegraphics[width=8cm]{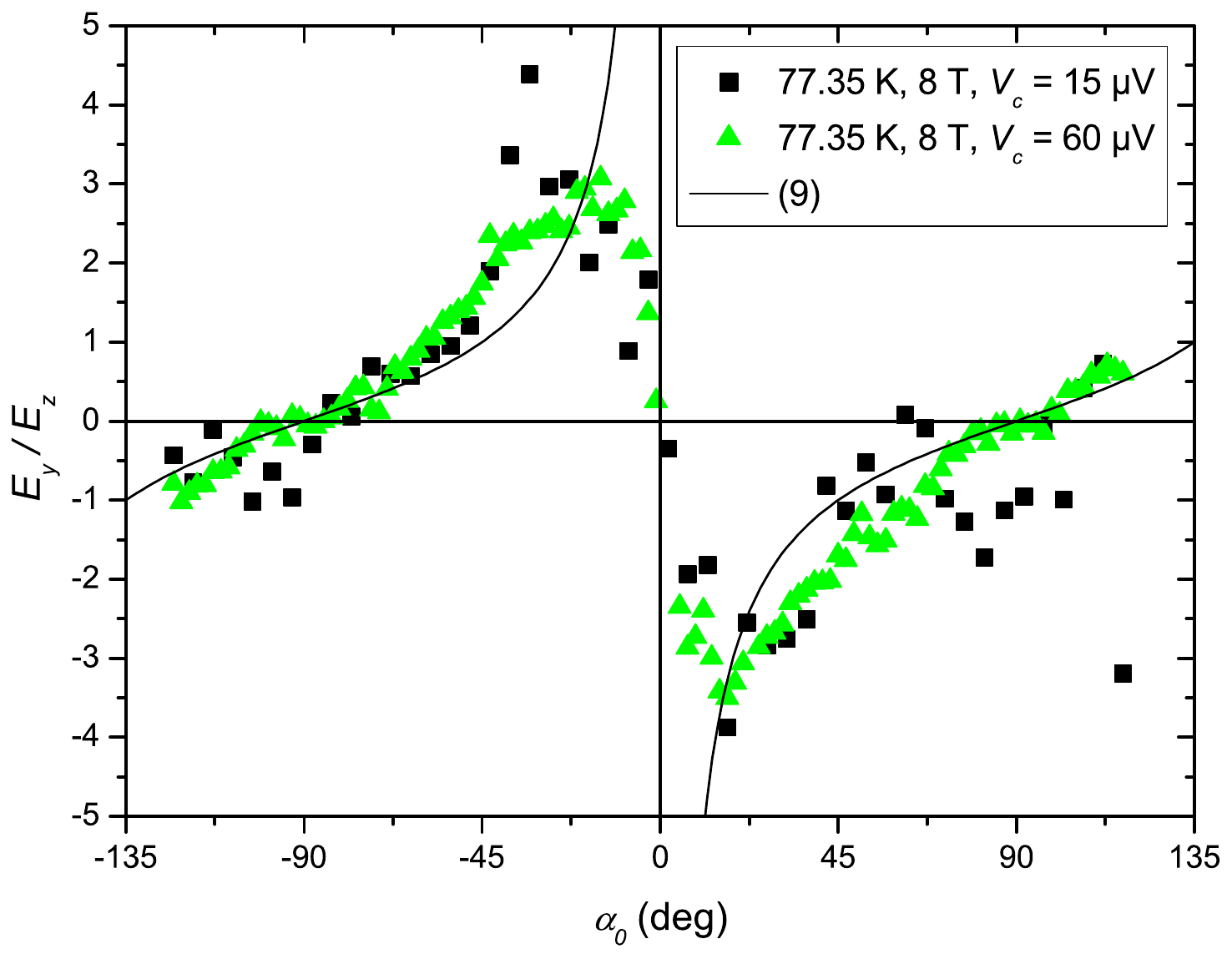}
\end{center}
\caption{The measured ratio $E_y/E_z$ vs $\alpha_0$ at $T = 77.35\K$ and $\mu_0H = 8\T$. The solid curve is  $E_y/E_z = -\cot\alpha_0$ (\ref{EybyEzTransport}), expected due to flux transport in the absence of flux-line cutting.}
\label{fig:Ez_CGHG_77p35K_8T}
\end{figure}

Similar measurements were also performed at  $T = 77.35\K$ and $\mu_0H = 8\T$ for $V_c = 15\uV$ (square symbols) and $60\uV$ (triangles), and the results are shown in figure~\ref{fig:Ez_CGHG_77p35K_8T}.  Note that, as in figure \ref{fig:Ez_CGHG_85K_1T}, the scatter in the values of $E_y/E_z$ becomes smaller  with increasing values of the voltage criterion $V_c$.

The data presented in Figs.\ \ref{fig:Ez_CGHG_85K_1T} and \ref{fig:Ez_CGHG_77p35K_8T}  show results measured at constant $E_z$, but we also did an analysis making use of the saved $V(I)$ data of both voltmeter channels to construct corresponding plots of $E_y/E_z$ vs $\alpha_0$ at constant $E_{tot} = \sqrt{E_y^2 + E_z^2}$.   While there were some differences from the results for constant $E_z$ for the smallest values of $V_c$, these were not significant for the larger values of $V_c$. 

\section{Comparison with theoretical predictions\label{Expt&Theory}}

\begin{figure}
\begin{center}
\includegraphics[width=8cm]{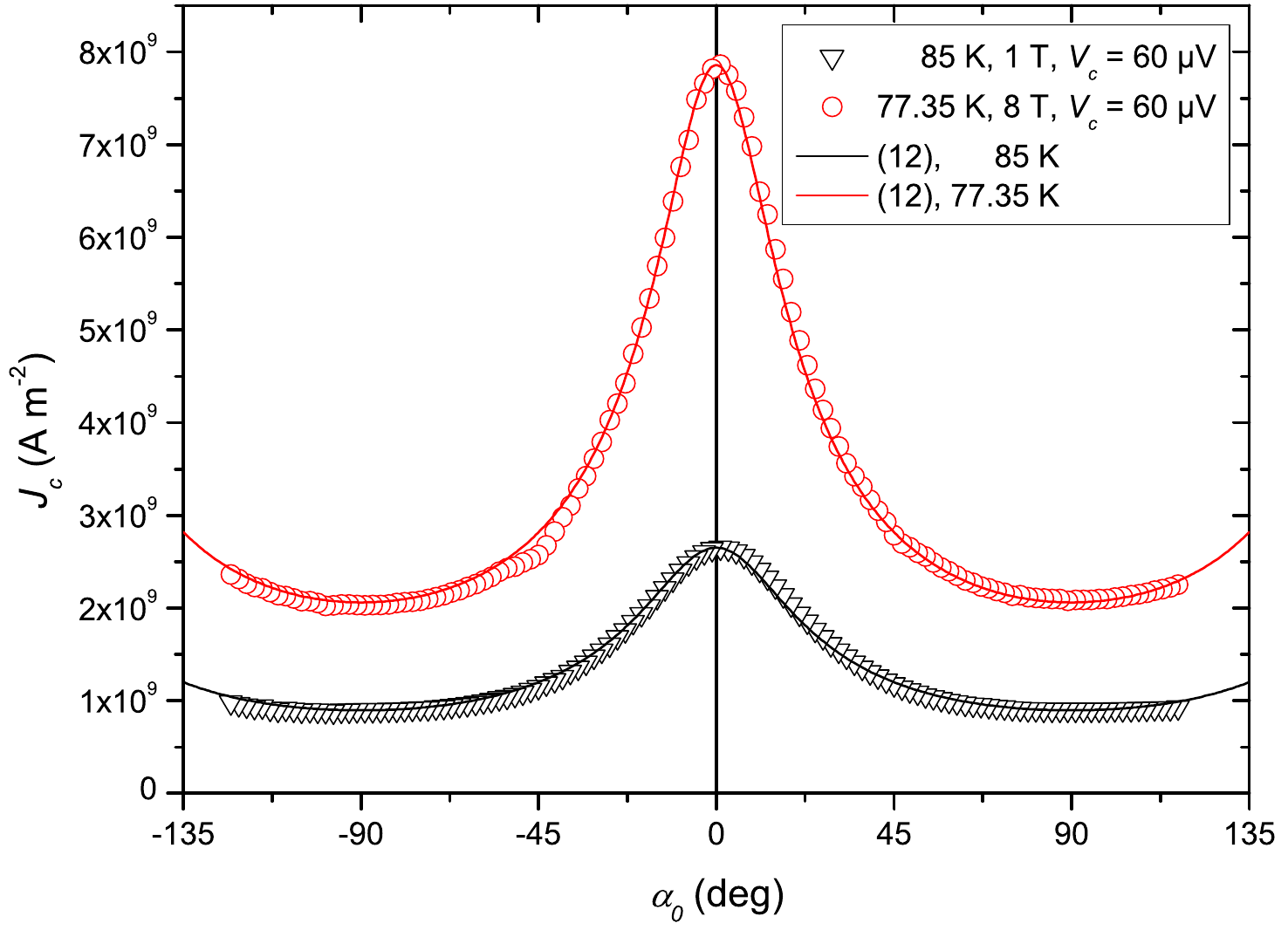}
\end{center}
\caption{Critical-current density $J_c$ ($V_c = 60\uV$, $E_z = 20$ mV/m) vs the angle $\alpha_0$ of the in-plane applied field at 85 K and 1 T (inverted triangles) and 77.35 K and 8 T (open circles).  The corresponding solid curves are fits using the elliptic model (\ref{Jcellipse}) with $J_{c\parallel} = 2.65 \times 10^9$ A/m$^2$ and $J_{c\perp} = 0.895 \times 10^9$ A/m$^2$ at 85 K and  $J_{c\parallel} = 7.86 \times 10^9$ A/m$^2$ and $J_{c\perp} = 2.06\times 10^9$ A/m$^2$ at 77.35 K.}
\label{Jcplots}
\end{figure}

Shown in figure \ref{Jcplots} are the critical current densities $J_c$ vs $\alpha_0$ measured at $T = 85\K$ and $\mu_0H = 1\T$ (inverted triangles) and  at $T = 77.35\K$ and $\mu_0H = 8\T$ (open circles), both with the voltage criterion $V_c = 60\uV$, which corresponds to $E_z = 20$ mV/m.  The corresponding solid curves are fits to the data using the elliptic critical-state model: at  $T = 85\K$ and $\mu_0H = 1\T$, $J_{c\parallel} = 2.65 \times 10^9$ A/m$^2$ and $J_{c\perp} = 0.895 \times 10^9$ A/m$^2$ (the average of values at $\pm 90\degree$), such that $J_{c\perp}/J_{c\parallel} = 0.338\pm 0.002$, and   at $T = 77.35\K$ and $\mu_0H = 8\T$,  $J_{c\parallel} = 7.86 \times 10^9$ A/m$^2$ and $J_{c\perp} = 2.06 \times 10^9$ A/m$^2$ (the average of values at $\pm 90\degree$), such that $J_{c\perp}/J_{c\parallel} = 0.262\pm 0.004$.  (The indicated errors are chosen such that both values lie within the range.)  The data are rather well fitted by the elliptic critical-state model as in  \cite{Herzog97}.  In agreement with previous experiments \cite{Durrell03,Rutter05,Durrell07,Maiorov07} the data clearly do not show the cusp-like maxima predicted by the GDCSM (see figure \ref{fig3}).  Thus the $J_c$ vs $\alpha_0$ measurements support  either the original elliptical critical-state model (OECSM) or the extended elliptical critical-state model (EECSM), since both assume the same dependence of $J_c(\phi)$, but do not support the generalized double critical-state model (GDCSM).

\begin{figure}
\begin{center}
\includegraphics[width=8cm]{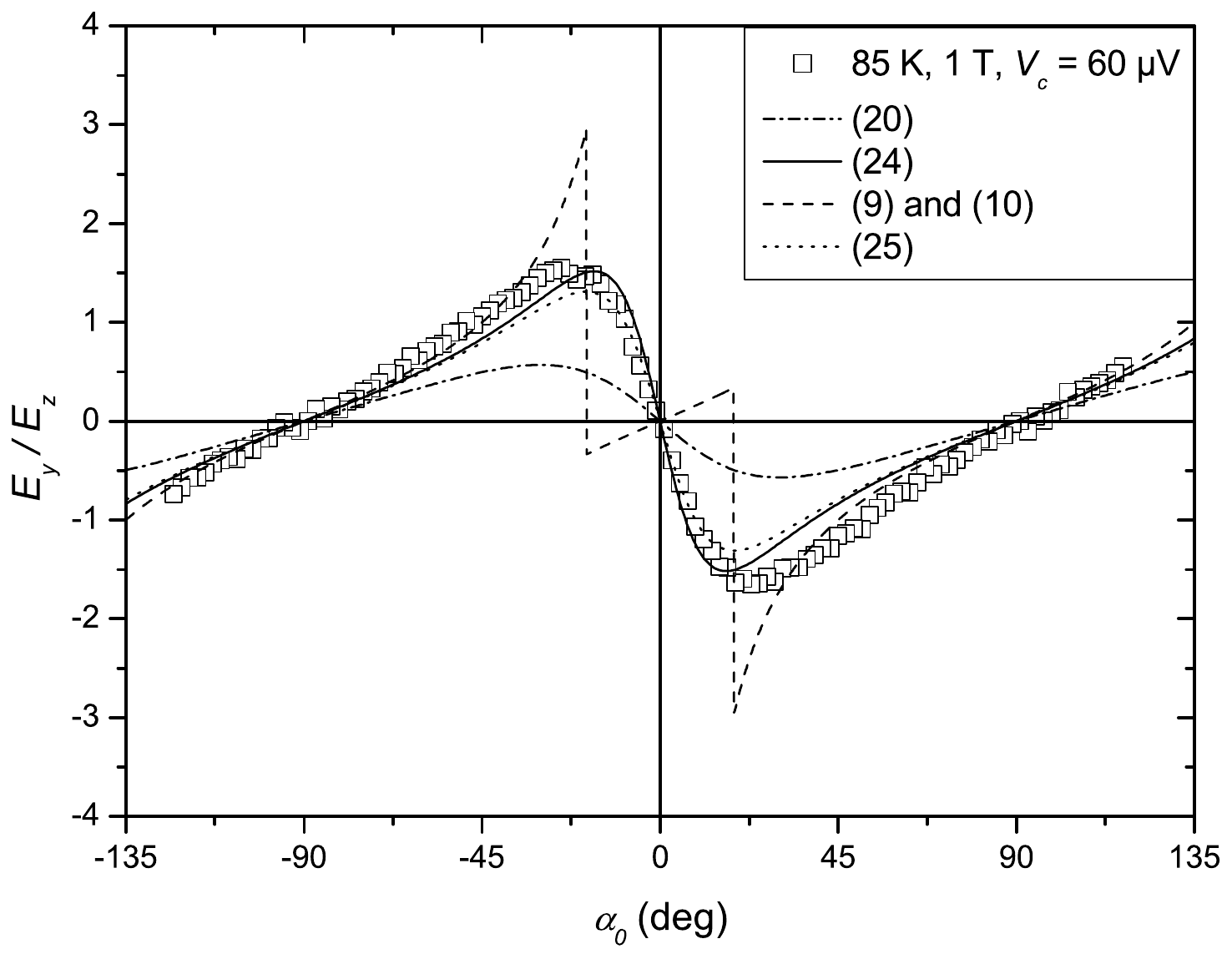}
\end{center}
\caption{Comparison of the measured  $E_y/E_z$ (open symbols) vs $\alpha_0$ at 85 K and 1 T ($V_c = 60\uV$) with theoretical curves for the OECSM  [(\ref{tanpsiE}), dot-dashed] and the BLRM [dotted, (\ref{BLRmodel})] with  $\tan\phi_c=J_{c\perp}/J_{c\parallel} = 0.338$,  the EECSM with  $r= \rho_c/\rho_f=0.090$ [(\ref{tanpsiEcreepa}), solid], and the GDCSM and BMM with  $\tan\phi_c=J_{c\perp}/J_{c\parallel} = 0.338$ [(\ref{EybyEzTransport}) and (\ref{EybyEzCutting}), dashed]. }
\label{EybyEz85Kplot}
\end{figure}

Figure \ref{EybyEz85Kplot} shows comparisons of the measurements of $E_y/E_z$ vs $\alpha_0$ at 85 K and 1 T (using  the voltage criterion $V_c = 60\uV$) with the predictions of the original elliptic critical-state model [dot-dashed curve,  (\ref{tanpsiE})] and the Bad\'ia-Maj\'os-L\'opez-Ruiz model [dotted curve, (\ref{BLRmodel})]
using the measured $\tan\phi_c=J_{c\perp}/J_{c\parallel} = 0.338$ , the extended  elliptic critical-state model using $r= \rho_c/\rho_f = 0.090$ assumed to be independent of $\alpha_0$  [solid curve,  (\ref{tanpsiEcreepa})], and the generalized double critical-state model and the Brandt-Mikitik model using the measured $\tan\phi_c=J_{c\perp}/J_{c\parallel} = 0.338$ [dashed curve, (\ref{EybyEzTransport}) and  (\ref{EybyEzCutting})].  The theoretical curves for the extended elliptic critical-state model and the Bad\'ia-Maj\'os-L\'opez-Ruiz model can account for the large values for the magnitude of the measured $E_y/E_z$, while the original elliptic critical-state model cannot. The data also clearly do not show the sawtoothlike features predicted by the  generalized double critical-state model and the Brandt-Mikitik model. To summarize, these measurements of $E_y/E_z$ vs $\alpha_0$ favor the extended elliptical critical-state model and the Bad\'ia-Maj\'os-L\'opez-Ruiz model but do not support the original elliptic critical-state model, the generalized double critical-state model, or the Brandt-Mikitik model.

\begin{figure}
\begin{center}
\includegraphics[width=8cm]{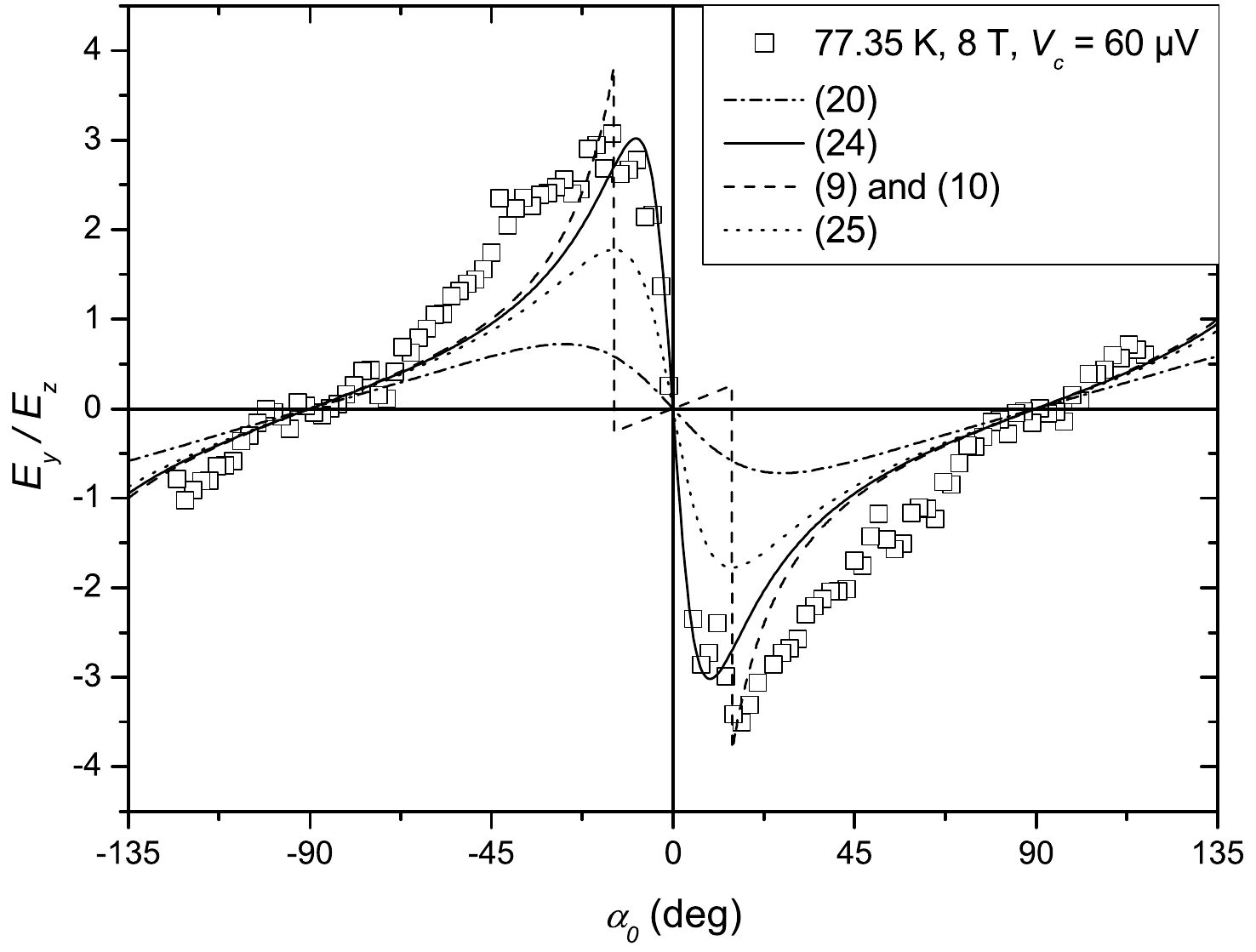}
\end{center}
\caption{Comparison of the measured  $E_y/E_z$ (open symbols) vs $\alpha_0$ at 77.35 K and 8 T ($V_c = 60\uV$) with theoretical curves for the OECSM  [(\ref{tanpsiE}), dot-dashed] and the BLRM [dotted, (\ref{BLRmodel})] with  $\tan\phi_c=J_{c\perp}/J_{c\parallel} = 0.262$,  the EECSM with  $r= \rho_c/\rho_f=0.026$ [(\ref{tanpsiEcreepa}), solid], and  the GDCSM and BMM with  $\tan\phi_c=J_{c\perp}/J_{c\parallel} = 0.262$ [(\ref{EybyEzTransport}) and (\ref{EybyEzCutting}), dashed]. }
\label{EybyEz77Kplot}
\end{figure}

Figure \ref{EybyEz77Kplot} shows comparisons of the measurements of $E_y/E_z$ vs $\alpha_0$ at  $T = 77.35\K$ and $8\T$ (using  the voltage criterion $V_c = 60\uV$) with the predictions of the original elliptic critical-state model  [dot-dashed curve,  (\ref{tanpsiE})] and  the Bad\'ia-Maj\'os-L\'opez-Ruiz model [dotted curve, (\ref{BLRmodel})]
using the measured $\tan\phi_c=J_{c\perp}/J_{c\parallel} = 0.262$, the extended  elliptic critical-state model using $r= \rho_c/\rho_f = 0.026$ assumed to be independent of $\alpha_0$  [solid curve,  (\ref{tanpsiEcreepa})], and the generalized double critical-state model and the Brandt-Mikitik model using the measured $\tan\phi_c=J_{c\perp}/J_{c\parallel} = 0.262$ [dashed curve, (\ref{EybyEzTransport}) and  (\ref{EybyEzCutting})]. The theoretical curve for the extended elliptic critical-state model can partially account for the large values for the magnitude of the measured $E_y/E_z$, but the original elliptic critical-state model and the Bad\'ia-Maj\'os-L\'opez-Ruiz model cannot.  The large magnitudes of the experimental values of $E_y/E_z$ at 77.35 K and 8 T cannot be understood with any of the proposed models without modification.

\begin{figure}
\begin{center}
\includegraphics[width=8cm]{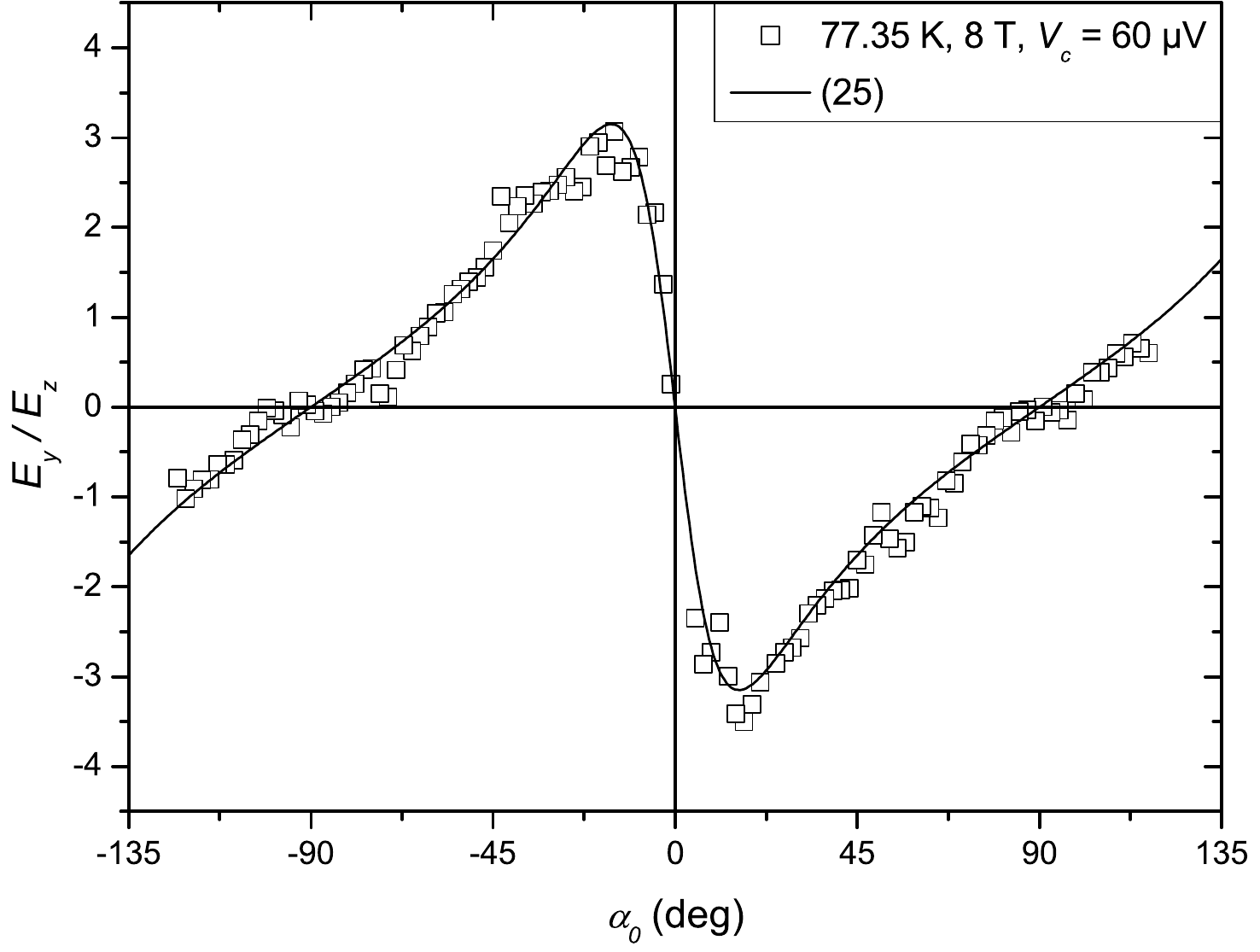}
\end{center}
\caption{The measured $E_y/E_z$ (open symbols) vs $\alpha_0$ at $77.35\K$ and $8\T$ ($V_c = 60 \uV$, $E_z = 0.02\unit{V/m}$) and the assumed fit function [(\ref{eq:ClemEq54prefactor}), solid curve] with the fit parameters $r= \rho_c/\rho_f = 0.080$ and $c = 1.936$)}
\label{fig:Ez_CGHG_77p35K_8T_fit_theory}
\end{figure}

Although we cannot at present provide a justification for the following assumption, if we assume the following form for theoretical values of $E_y/E_z$, 
\begin{equation}
\frac{E_y}{E_z} = c \frac{(r-1) \tan \alpha_0}{r + \tan^2 \alpha_0},
\label{eq:ClemEq54prefactor}
\end{equation}
we obtain a considerably better fit to the data at 77.35 K and 8 T, as shown in figure \ref{fig:Ez_CGHG_77p35K_8T_fit_theory}, where the fit parameters were found to be $r = 0.080$ and $c = 1.936$.

\section{\label{Conclusions}%
Conclusions}

The purpose of this paper has been to compare five theoretical models describing the critical state of type-II superconductors when the current density $\bm J$ is at an arbitrary angle with respect to the local magnetic flux density $\bm B$ and then to use experimental measurements of the angular dependence of the critical current density $J_c$ and the electric field $\bm E$ (for $J$ just above $J_c$) as a test of these theories.  

For simplicity, the geometry of an infinite, macroscopically isotropic type-II superconducting slab of finite thickness centered on the $yz$ plane was used for calculating the predictions of the five  theories.  However, the experiments were performed using a thin YBCO film subjected to in-plane currents and magnetic fields, but we assumed that the effects of finite dimensions and self-fields were small enough that the experiments would be able to make a clear distinction among the five theories.

Measurements of the angular dependence of the critical-current density $J_c$  showed that the cusp-like behaviour predicted by the generalized double critical-state model (GDCSM) was not verified.  Instead, the angular dependence of $J_c$ showed a behaviour very similar to that assumed by both the original elliptical critical-state model (OECSM) and the extended elliptical critical-state model (EECSM).  

Measurements of the angular dependence of the ratio of the transverse to the longitudinal components of the electric field $E_y/E_z$  at 85 K and 1 T for $J$ just above $J_c$ also showed that the sawtoothlike behaviour predicted by the generalized double critical-state model (GDCSM) and the Brandt-Mikitik model (BMM) was not verified. While all three elliptical critical-state models predict a  smooth angular dependence of $E_y/E_z$, as found experimentally, the magnitude of  $E_y/E_z$ predicted by the original elliptical critical-state model (OECSM) was  too small to account for the experimental values.  On the other hand, both the extended  elliptical critical-state model (EECSM) with  $r= \rho_c/\rho_f = 0.090$ and the Bad\'ia-Maj\'os-L\'opez-Ruiz model (BLRM)  were found to be in reasonable agreement with the data at 85 K and 1 T.  However, the EECSM has the advantage that it is applicable for currents $J$ well above $J_c(\phi)$, while the BLRM is currently limited to $J$ just above $J_c(\phi)$.

None of the five models was found to provide a good description of the data for $E_y/E_z$ at 77.35 K and 8 T.  A small value of $r= \rho_c/\rho_f = 0.026$ within the extended  elliptical critical-state model (EECSM) can account roughly for the large magnitudes of $E_y/E_z$, but the resulting theoretical curve does  not provide a good fit to the shape of the data.  By adding an additional fit parameter to the theoretical expression for $E_y/E_z$ within the EECSM, better agreement with the magnitude and shape of the data for $E_y/E_z$ at 77.35 K and 8 T can be obtained, but at present we have no theoretical justification for this assumption.  Further theoretical and experimental work to investigate the influence of  finite dimensions and self-fields will be needed to shed light on this problem.  Our theory assumes homogeneity of the electric field $\bm E$, but this needs to be confirmed, since earlier experiments measuring the local electric field in the longitudinal geometry found strong inhomogeneities in $\bm E $ along the sample length \cite{Irie74,Irie75,Ezaki76,Cave78,Matsushita98}.

The experimental results in this paper call attention to two important features that to date have not been generally recognized.  First, when $\bm J$ is neither parallel nor perpendicular to the local magnetic flux density $\bm B$, both flux cutting and flux transport occur simultaneously when $J$ exceeds the critical current density $J_c$.  This indicates the intimate relationship between flux cutting and depinning.  Second, the dynamical properties of the superconductor when $J$ exceeds $J_c$ depend in detail upon two effective resistivities, $\rho_c$ and $\rho_f$, and their ratio $r= \rho_c/\rho_f$.  The consequence of this is that to provide a full description of the properties of type-II superconductors in the resistive state, both $\rho_c$ and $\rho_f$ should be measured.  It can be shown that in cylindrical geometry, values of $r < 1$ are responsible for the paramagnetic moment generated along the length of a current-carrying type-II superconductor above its critical current when it is subjected to a longitudinal applied magnetic field \cite{Walmsley72a}.

It seems likely that the angular dependence of the critical current density in type-II superconductors will depend upon the details of the underlying pinning structure, and it is possible that in many cases $J_c(\phi)$, which can be obtained experimentally,   will not be given precisely by the elliptic form given in  (\ref{Jcellipse1}) and (\ref{Jcellipse}).  For example, $J_c(\phi)$ may even take the form of a superellipse, as suggested in  \cite{Badia09} and \cite{Ruiz11}.  In other words, although one should not take the elliptic model too seriously, it seems likely that $J_c(\phi)$ generally has roughly the shape shown in figure \ref{fig5}, and that it is a general principle that both flux cutting and flux transport occur simultaneously when $J > J_c(\phi)$.

\ack{We thank  V G Kogan for
stimulating discussions. This research, supported by the U.S. Department of
Energy, Office of Basic Energy Science, Division of Materials
Sciences and Engineering, was performed in part at
the Ames Laboratory, which is operated for the U.S. Department
of Energy by Iowa State University under Contract No.
DE-AC02-07CH11358. This work was supported by the Engineering and Physical Sciences Research Council (grant number EP/C011546/1) and by the PhD Plus scheme.}

\end{document}